\documentclass[preprint,showpacs,preprintnumbers,amsmath,amssymb,nofootinbib,superscriptaddress]{revtex4-1}

\usepackage{graphicx}
\usepackage{float}
\usepackage{wrapfig}
\usepackage{subcaption}
\usepackage[justification=justified, font=footnotesize, width=0.8\textwidth]{caption}
\usepackage{tabularx}
\usepackage{relsize}
\usepackage{color}
\usepackage[normalem]{ulem}

\usepackage{bm}

\usepackage{hyperref}

\usepackage{verbatim}
\usepackage{amsmath,amssymb}

\begin{document}
	
\title{The measure of PBR's reality}

\author{S\'anchez-Kuntz, Natalia}
\affiliation{Institut f\"ur Theoretische Physik Universit\"at Heidelberg\\ 
Philosophenweg 16, D-69120 Heidelberg\footnote[0]{\href{mailto:sanchez@thphys.uni-heidelberg.de}{sanchez@thphys.uni-heidelberg.de}}}
\author{Nahmad-Achar, Eduardo}
\affiliation{Instituto de Ciencias Nucleares Universidad Nacional Aut\'onoma de M\'exico Apdo. Postal 70-543 M\'exico, Cd.Mx., 04510\footnote[0]{\href{mailto:nahmad@nucleares.unam.mx}{nahmad@nucleares.unam.mx}}}

\begin{abstract}
\noindent We review the Pusey-Barret-Rudolph (PBR) theorem\cite{PBR} and their setup, and arrive to the conclusion that the reality of a quantum state $\psi$ is intrinsically attached to the measurement the system described by $\psi$ has undergone. We show that a state that has not been measured can be regarded as pure information, while a state that has been measured has to be regarded as a physical property of a certain system, having a counterpart in reality. This demonstration implies that the statement of PBR's theorem changes in a meaningful way. 
\end{abstract}


\maketitle

\section{Theoretical framework, the PBR paper}

Paradoxical aspects of Quantum Mechanics (QM) and perplexing behaviours of matter have been shown to everyone who has ever come close to the theory.

A comforting way of viewing some of the predictions of QM consists in letting all the mathematics be only descriptive of the \emph{real} behaviour of quantum systems. In that case, one could think that the wave function is only a mathematical tool, and there is nothing ``wavy'' about an electron, except that it behaves according to probabilities established by its wave function.

In a recent paper\cite{PBR}, Pusey, Barrett and Rudolph (PBR) demonstrate that, assuming realism, the wave function cannot be interpreted in that manner, that is, that the wave function must be ontologically attached to the system it describes. The statement of their theorem reads:
\begin{quote}
[If] a system has a ‘real physical state’\textemdash not necessarily completely described by quantum theory, but objective and independent of the observer... [and if] systems that are prepared independently have independent physical states[, then a pure state cannot be regarded as mere information, or else one would arrive to a contradiction when faced with the quantum theoretical framework.]
\end{quote}

\begin{figure}[H]
\begin{center}
	\includegraphics[width=0.5\textwidth]{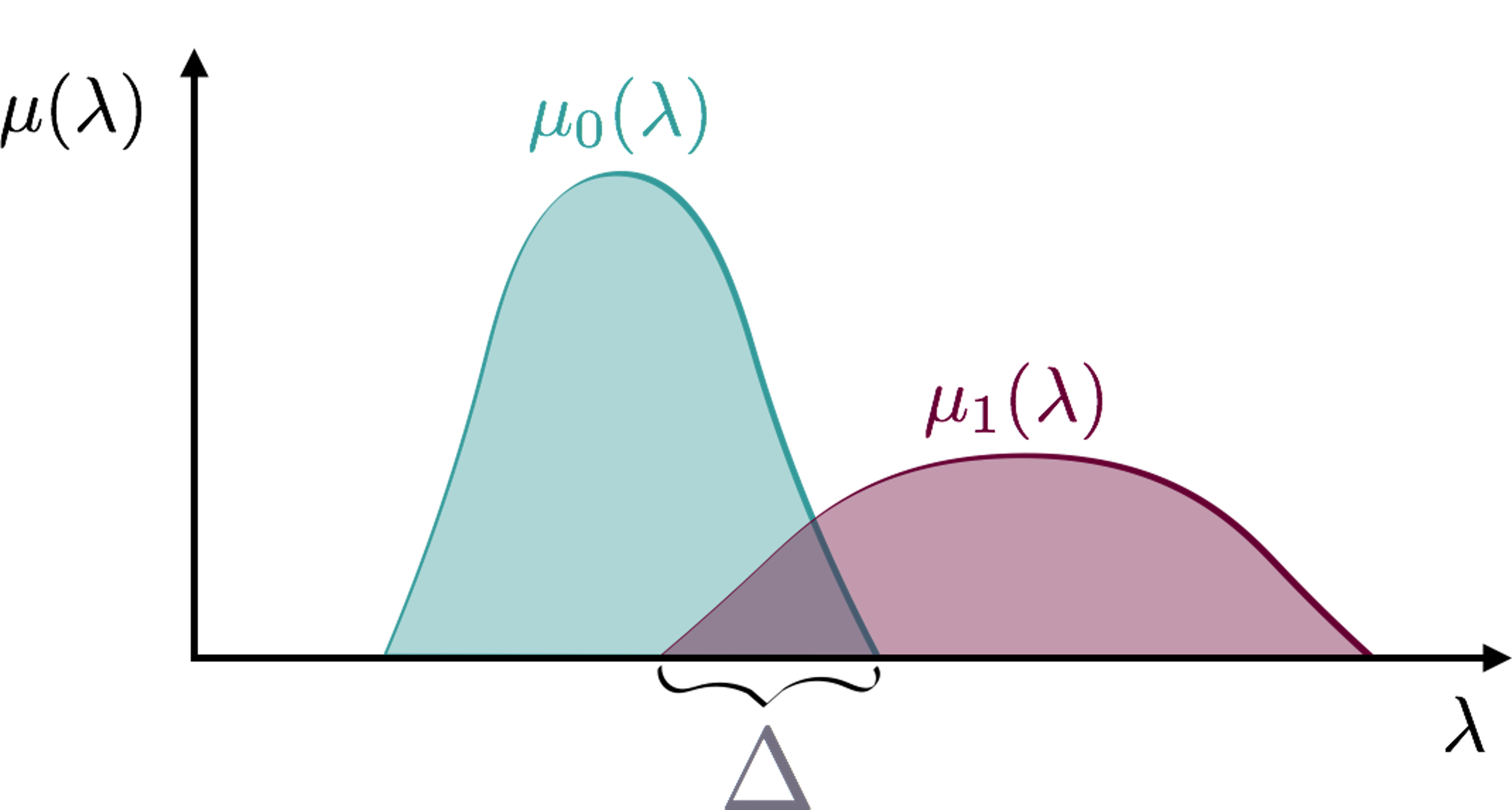}	
	\caption{The notion of $\vert \psi_0 \rangle$  and $\vert \psi_1 \rangle$ representing mere information is illustrated. $\vert \psi_0 \rangle$  and $\vert \psi_1 \rangle$ are two non-orthogonal states.  $\mu_0(\lambda)$ is the probability distribution of $\vert \psi_0 \rangle$ over the space of physical states, and $\mu_1(\lambda)$ is that of $\vert \psi_1 \rangle$. The defining quality of representing mere information is that these two distributions overlap, such that there is a non-zero probability that both preparations\textemdash $\vert \psi_0 \rangle$ and $\vert \psi_1 \rangle$\textemdash result in a physical state $\lambda_p \in \Delta$.} 
\label{gPBR}
\end{center}
\end{figure}

They carry out their demonstration by adopting the notion given by Harrigan and Spekkens\cite{HS} of what `mere information' looks like. This notion goes as follows: if $\lambda$ is a label for the physical state of the system, and two different, but non-orthogonal preparations, $\vert \psi_0 \rangle$ and $\vert \psi_1 \rangle$, are mere information about some physical state, then the distribution $\mu_0(\lambda)$, which assigns the probability of $\vert \psi_0 \rangle$ resulting in a physical state $\lambda$, \emph{and} the distribution $\mu_1(\lambda)$ of $\vert \psi_1 \rangle$, overlap; so that $\vert \psi_0 \rangle$ and $\vert \psi_1 \rangle$ can both result in a physical state $\lambda_p$ from the overlap region with non-zero probability. Conversely, if the distributions  $\mu_0(\lambda)$ and $\mu_1(\lambda)$ overlap, one can affirm that $\vert \psi_0 \rangle$ and $\vert \psi_1 \rangle$ represent mere information about the underlying physical state of the system. The scheme of this notion is given in Fig. \ref{gPBR}.

PBR go on to consider two identical and independent preparation devices; each device prepares a system in either the quantum state $\vert \psi_0 \rangle = \vert 0 \rangle$ or the quantum state 
$\vert \psi_1 \rangle = \vert + \rangle = (\vert 0 \rangle + \vert 1 \rangle)/\sqrt{2}$, so that when the two states are brought together, the complete system is compatible with any of the four quantum states: 
\begin{equation}
	\label{qstates}
\vert 0 \rangle \otimes \vert 0 \rangle, \hspace{5pt} \vert 0 \rangle \otimes \vert + \rangle, \hspace{5pt} \vert + \rangle \otimes \vert 0 \rangle, \hspace{5pt} \mbox{and} \hspace{5pt} \vert + \rangle \otimes \vert + \rangle.
\end{equation}
Then, they assume that if the two states $\vert \psi_0 \rangle$ and $\vert \psi_1 \rangle$ represent mere information, there is a probability $q^2 > 0$ that both systems result in physical states, $\lambda_1$ and $\lambda_2$, from the overlap region, $\Delta$ (cf. Fig. \ref{gPBR}). 
The complete system can be measured, and for this they propose an entangled measurement with the four possible outcomes: 
$$
\vert \xi_1 \rangle =  \frac{1}{\sqrt{2}} \Big[\vert 0 \rangle \otimes \vert 1 \rangle \, + \, \vert 1 \rangle \otimes \vert 0 \rangle \Big],
$$
$$
\vert \xi_2 \rangle =  \frac{1}{\sqrt{2}} \Big[\vert 0 \rangle \otimes \vert - \rangle \, + \, \vert 1 \rangle \otimes \vert + \rangle \Big],
$$
$$
\vert \xi_3 \rangle =  \frac{1}{\sqrt{2}} \Big[\vert + \rangle \otimes \vert 1 \rangle \, + \, \vert - \rangle \otimes \vert 0 \rangle \Big],
$$
$$
\vert \xi_4 \rangle =  \frac{1}{\sqrt{2}} \Big[\vert + \rangle \otimes \vert - \rangle \, + \, \vert - \rangle \otimes \vert + \rangle \Big];
$$
but the probability that the quantum state $\vert 0 \rangle \otimes \vert 0 \rangle$ results in $\vert \xi_1 \rangle$ is zero, same for $\vert 0 \rangle \otimes \vert + \rangle$ resulting in $\vert \xi_2 \rangle$, for $\vert + \rangle \otimes \vert 0 \rangle$ resulting in $\vert \xi_3 \rangle$, and for $\vert + \rangle \otimes \vert + \rangle$ resulting in $\vert \xi_4 \rangle$. This takes them to the conclusion that if the state $\lambda_1 \otimes \lambda_2$ that arrives to the detector is compatible with the four quantum states (\ref{qstates}), then the measuring device could give a result that should, following simple QM, occur with zero probability. This contradiction arises only by assuming that the distributions of $\vert 0 \rangle$ and $\vert + \rangle$ overlap, so their distributions cannot overlap, and therefore $\vert \psi_0 \rangle = \vert 0 \rangle$ and $\vert \psi_1 \rangle =\vert + \rangle$ cannot represent mere information of an underlying physical system.

They extend their demonstration to any pair of quantum states, 
$$
\vert \psi_0 \rangle =  \cos\frac{\theta}{2} \vert 0 \rangle + \sin\frac{\theta}{2} \vert 1 \rangle,
$$
$$
\vert \psi_1 \rangle =  \cos\frac{\theta}{2} \vert 0 \rangle - \sin\frac{\theta}{2} \vert 1 \rangle,
$$
and thereby show that pure states must have a direct counterpart in reality: that either $\vert \psi \rangle$ must represent a physical property of the physical system, or that realism cannot be an ingredient of QM, or even that system independency is violated.

\section{What is a state without a measurement?}

In this section we will demonstrate that that the no-go theorem for the reality of the quantum state given by PBR necessarily implies a measurement over the quantum system at the preparation stage, whose state turns out to be real.

Suppose that we have two identical preparation devices each of which prepares either the state $ \vert 0 \rangle$ or the state $\vert + \rangle$ \emph{without making an announcement of which of the two states was prepared}. 

\begin{figure}[H]
\begin{center}
	\includegraphics[width=0.7\textwidth]{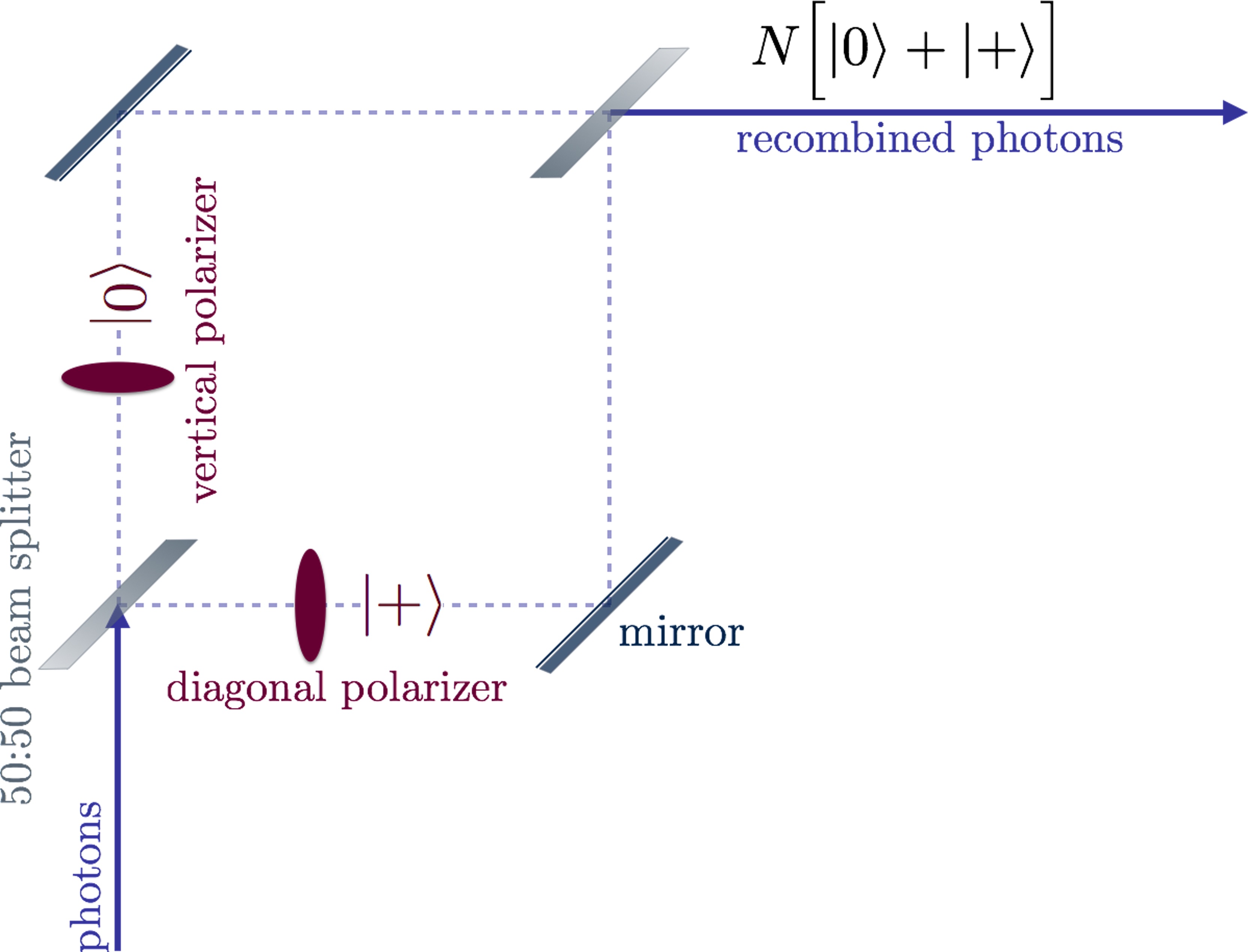}	
	\caption{Preparation device of the states $\vert 0 \rangle$ and $\vert + \rangle$ with no distinction between them.} 
	\label{recombined}
\end{center}
\end{figure}

Each device can be built by putting together a photon source and a Mach-Zehnder interferometer, with a polarizer set in the vertical direction $(\equiv \vert 0 \rangle)$ on one path of the laser beam, and a polarizer set to $45^\circ$ $(\equiv \vert + \rangle)$ on the other path of the laser beam. At the end, the two paths are recombined (cf. Fig. \ref{recombined}).

\begin{figure}[H]
\begin{center}
\begin{subfigure}{0.4\textwidth}
\begin{center}
\includegraphics[width=0.8\linewidth]{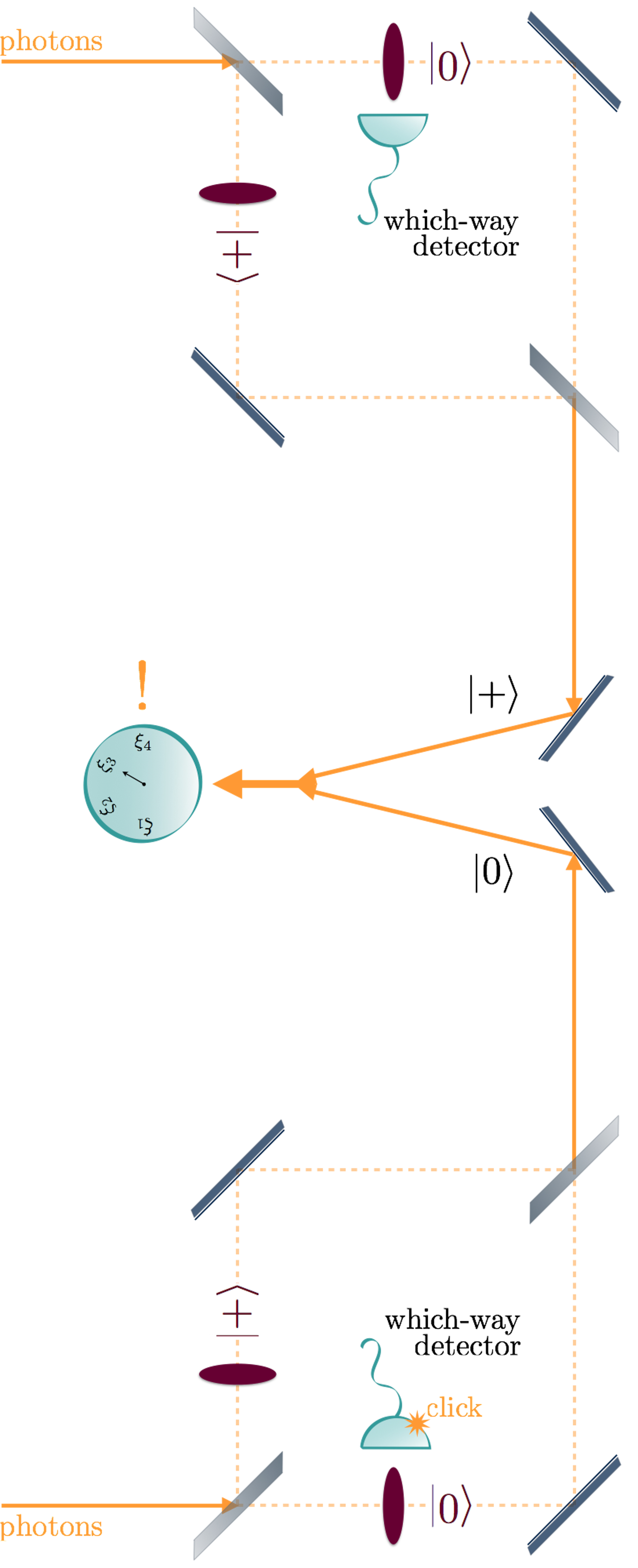} 
\end{center}
\caption{The contradiction derived by\\ PBR arises.}
\label{setupa}
\end{subfigure}
\begin{subfigure}{0.4\textwidth}
\begin{center}
\includegraphics[width=0.8\linewidth]{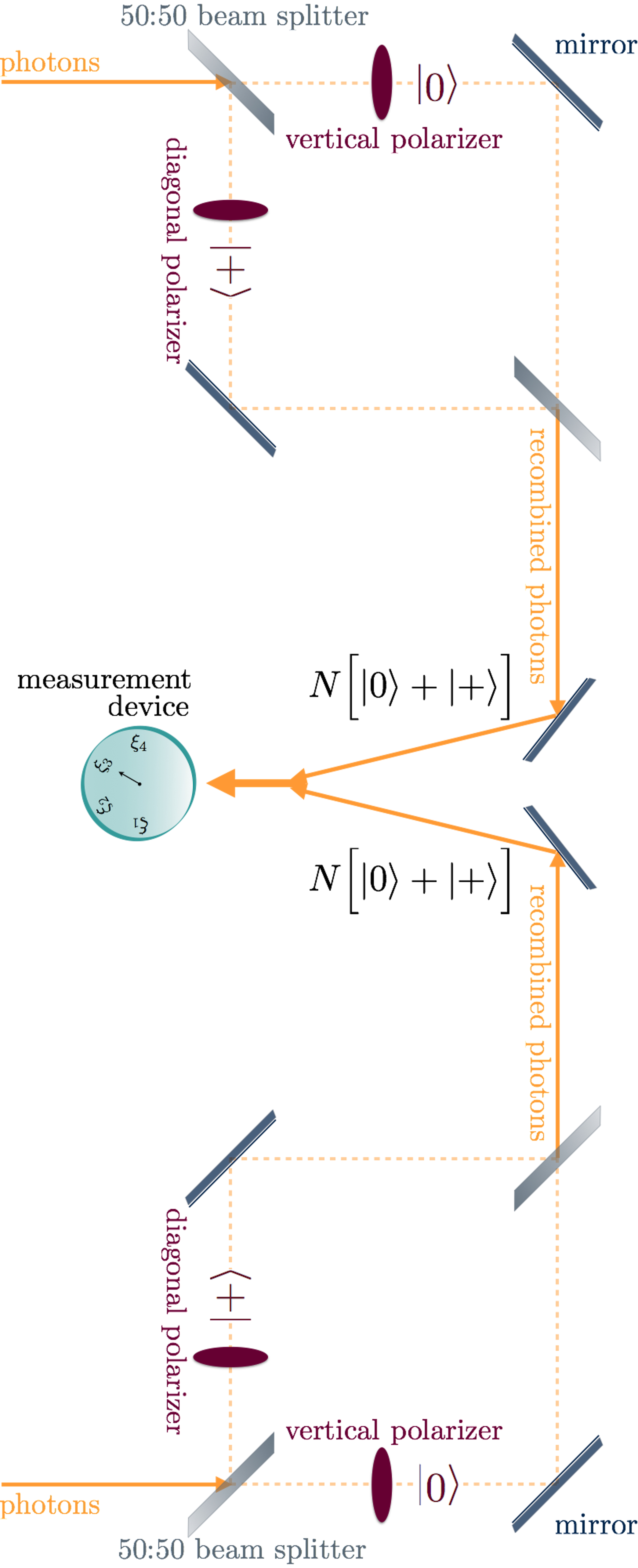} 
\end{center}
\caption{No contradiction arises.}
\label{setupb}
\end{subfigure}
\caption{Each experimental setup is a combination of two identical and independent preparation devices. In \textbf{(a)} the preparation devices consist of a source of photons which undergoes a splitting of half and half in two optical paths, each optical path is polarized in a certain direction and the two paths are recombined. Moreover, two which-way detectors are added, one for each preparation device: if a detector clicks, the photon that has been prepared is in the state $\vert 0 \rangle$, while if a detector does not click, the photon comes out in the state $\vert + \rangle$. At the end, the photons of the two preparation devices are brought together and measured. In \textbf{(b)} the preparation devices are as those of (a), but without the two which-way detectors. The photon coming out of each of the two paths and into the measurement device is therefore in the state $N [\vert 0 \rangle + \vert + \rangle]$.}
\label{setups}
\end{center}
\end{figure}

Under these conditions, the state that comes out of the preparation device is
$$
\vert \psi \rangle=  N \Big[\vert 0 \rangle + \vert + \rangle \Big].
$$
One can easily calculate the normalization constant $N$ to obtain
$$
N^2=\frac{\sqrt{2}}{2\sqrt{2}+2}\,.
$$
Finally, when two identical preparation devices are put together (cf. Fig. \ref{setups}), the state that arrives at the detector is
$$ 
\vert \Psi \rangle = \vert \psi \rangle \otimes \vert \psi \rangle = N^2 \Big[\vert 0 \rangle + \vert + \rangle \Big] \otimes \Big[\vert 0 \rangle + \vert + \rangle \Big],
$$

\noindent (see Fig. \ref{setupb}), and not one of the states (\ref{qstates}) assumed by PBR. This state $\vert \Psi \rangle$ that arrives at the detector is compatible with the measurement basis used in the PBR theorem, 
$$
\vert \xi_1 \rangle =  \frac{1}{\sqrt{2}} \Big[\vert 0 \rangle \otimes \vert 1 \rangle \, + \, \vert 1 \rangle \otimes \vert 0 \rangle \Big],
$$
$$
\vert \xi_2 \rangle =  \frac{1}{\sqrt{2}} \Big[\vert 0 \rangle \otimes \vert - \rangle \, + \, \vert 1 \rangle \otimes \vert + \rangle \Big],
$$
$$
\vert \xi_3 \rangle =  \frac{1}{\sqrt{2}} \Big[\vert + \rangle \otimes \vert 1 \rangle \, + \, \vert - \rangle \otimes \vert 0 \rangle \Big],
$$
$$
\vert \xi_4 \rangle =  \frac{1}{\sqrt{2}} \Big[\vert + \rangle \otimes \vert - \rangle \, + \, \vert - \rangle \otimes \vert + \rangle \Big],
$$
in the sense that it may result in any of its elements with non-zero probability; following the logic of PBR, no contradiction arises when regarding $\vert 0 \rangle$ and $\vert + \rangle$ as mere information. 

What this proves is that, \textbf{if there is no distinguishability in the preparation devices that PBR construct, and both are independent, the states} $\bm{\vert 0 \rangle}$ \textbf{and} $\bm{\vert + \rangle}$ \textbf{\emph{can} be regarded as mere information}; i.e., they could both span physical states in the overlapping region $\Delta$ of phase space (cf. Fig. \ref{gPBR}), so that the system that would arrive at the detector would be compatible with any of the four quantum states $\vert 0 \rangle \otimes \vert 0 \rangle$, $\vert 0 \rangle \otimes \vert + \rangle$, $\vert + \rangle \otimes \vert 0 \rangle$, and $\vert + \rangle \otimes \vert + \rangle$, \emph{and} compatible with any of the four measurement results, $\vert \xi_1 \rangle$, $\vert \xi_2 \rangle$, $\vert \xi_3 \rangle$, and $\vert \xi_4 \rangle$.

On the other hand, if we have a preparation procedure which distinguishes the outcomes $\vert 0 \rangle$ and $\vert + \rangle$ between them, as must be the case for the PBR theorem (cf. Fig. \ref{setupa}), then the system that arrives to the detector is in fact in one of the four states $\vert 0 \rangle \otimes \vert 0 \rangle$, $\vert 0 \rangle \otimes \vert + \rangle$, $\vert + \rangle \otimes \vert 0 \rangle$, and $\vert + \rangle \otimes \vert + \rangle$, and in such a case the states $\vert 0 \rangle$ and $\vert + \rangle$ cannot be regarded as mere information: they must arise from disjoint regions of phase space, and thus represent a physical property of the system.

Overall, we can conclude that the experimental setup PBR propose entails a preparation method that makes the state $\vert 0 \rangle$ distinguishable from the state $\vert + \rangle$ and thus gives the observer knowledge of the state in which each subsystem is being prepared. If there were no distinguishability between the two prepared subsystems, then, as has been shown, no contradiction would arise between regarding $\vert 0 \rangle$ and $\vert + \rangle$ as mere information and the measurement conditions imposed by PBR.

To stress the distinction between the two scenarios depicted in Fig. \ref{setups}, we add that the only difference between the two experimental setups (PBR's and ours) is that in the PBR experimental setup there must be a certain knowledge of the state of the system that is being prepared, and this knowledge entails a measurement of some kind, while in our experimental setup there is no measurement whatsoever between the preparation procedure and the detector. If we regard states $\vert 0 \rangle$ and $\vert + \rangle$ as mere information in the first experimental setup (\ref{setupa}), the contradiction derived by PBR arises, while if we regard $\vert 0 \rangle$ and $\vert + \rangle$ as mere information in the second experimental setup (\ref{setupb}), no contradiction arises.

We can conclude from this that a state that has not been measured (when prepared) can be regarded as pure information, while a state that has been measured has to be regarded as a physical property of a certain system, having a counterpart in reality.

\section{Conclusions}
We have made explicit that the preparation devices PBR propose only work as they claim if there is distinguishability between the preparation of the state $\vert 0 \rangle$ and the state $\vert + \rangle$. We know that distinguishability entails measurement. We have thus concluded that, in order to follow PBR's theorem to the consequence that a given quantum state shall not be regarded as mere information, the precise system which might be in such a quantum state must have undergone a certain measurement.

\section*{Acknowledgements}
\noindent
This work was partially supported by Direcci\'on General de Asuntos del Personal Acad\'emico, Universidad Nacional Aut\'onoma de M\'exico (under Project No.~IN101217). NS-K thanks Consejo Nacional de Ciencia y Tecnolog\'ia-M\'exico for financial support.


\end{document}